\pdfoutput=1

\documentclass[sigconf,nonacm]{aamas}  

\usepackage{balance} 
\usepackage{subcaption}
\usepackage{twoxtwogame}
\usepackage{multirow}
\usepackage{cprotect}
\usepackage{hyperref}

\setcopyright{none}  
\acmConference[arXiv]{arXiv}{May 2024}{}{}   
\copyrightyear{2024}
\acmYear{2024}
\acmDOI{}
\acmPrice{}
\acmISBN{}

\acmSubmissionID{<<EasyChair submission id>>}

\title[twoxtwogame LaTeX Package]{Visualizing 2×2 Normal-Form Games:\\twoxtwogame LaTeX Package}

\author{Luke Marris}
\affiliation{
  \institution{Google DeepMind}
  \country{United Kingdom}}
\email{marris@google.com}

\author{Ian Gemp}
\affiliation{
  \institution{Google DeepMind}
  \country{United Kingdom}}

\author{Siqi Liu}
\affiliation{
  \institution{Google DeepMind}
  \country{United Kingdom}}

\author{Joel Z. Leibo}
\affiliation{
  \institution{Google DeepMind}
  \country{United Kingdom}}

\author{Georgios Piliouras}
\affiliation{
  \institution{Google DeepMind}
  \country{United Kingdom}}

\begin{abstract}
Normal-form games with two players, each with two strategies, are the most studied class of games. These so-called 2×2 games are used to model a variety of strategic interactions. They appear in game theory, economics, and artificial intelligence research. However, there lacks tools for describing and visualizing such games. This work introduces a \LaTeX{} package for visualizing 2×2 games. This work has two goals: first, to provide high-quality tools and vector graphic visualizations, suitable for scientific publications. And second, to help promote standardization of names and representations of 2×2 games. The \LaTeX{} package, \texttt{twoxtwogame}, is maintained on GitHub and mirrored on CTAN, and is available under a permissive Apache 2 license.
\end{abstract}


\newcommand{\BibTeX}{\rm B\kern-.05em{\sc i\kern-.025em b}\kern-.08em\TeX}

\begin{document}


\pagestyle{fancy}
\fancyhead{}


\maketitle 


\section{Introduction}

Normal-form games (NFGs) are the simplest of the game representations, and 2×2 games are the simplest instantiations of NFGs. 2×2 games represent strategic interactions between two players, were each simultaneously selects one of two strategies, resulting in a payoff for each player. Despite their simplicity, such games can describe a rich variety of strategic interactions which may be: competitive (like zero-sum games), cooperative (like common-payoff games), or mixed-motive. Therefore it is not surprising they are used to model many real-world interactions.

2×2 games are ubiquitous, and their study \citep{harold2002_atlas_of_interpersonal_situations} is crucial to understanding cooperation \citep{gauthier1986_morals_by_agreement}, competition, coordination, conflictual coordination \citep{vanderschraaf2018strategic}, reciprocal altruism  \citep{wilkinson1984_vampire_bat}, incentive structures \citep{sugden1986_economics_of_rights_cooperation_and_welfare}, social dilemmas \citep{bruns2021_archetypal_games}, utilitarian behaviour, rational behaviour \citep{gintis2014_bounds_of_reason}, and seemingly irrational behaviour \citep{camerer1997progress} including framing effects \citep{de1997gain}. As a modeling paradigm, they focus on quantifiable incentives, which they regard as the ``rules of the game'', and thus the sole determinant of rational strategy. The space of 2×2 matrix games is also very rich and flexible, with many affordances for the modeler. Simply changing the outcomes associated to joint action can create very different strategic situations. The great flexibility and expressiveness of 2×2 games is why they have been used in so many different application domains, including sustainable economic policy modeling \citep{ostrom1994_rules}, social structures \citep{skyrms2004_stag_hunt,bicchieri2005_gammar_of_society,binmore1994_social_contract}, foreign policies \citep{schelling1966_arms_and_influence}, pandemic responses, and environmental treaties \citep{breton2006_environmental_projects,branzei2021_economies,schosser2022_fairness_pandemic}. 2×2 matrix games have also motivated many influential laboratory experiments on human behavior \citep{andreoni1993rational, cooper1996cooperation, schmidt2003playing} and individual differences \citep{sheldon1999learning}. Since experiments where human participants play matrix games have so few free parameters and are so easy to conduct, it is not hard to ensure that different labs can implement them in a consistent fashion. This feature allows them to be used across very different settings around the world, minimizing the effect of the different labs, and thereby making it possible to focus on cross cultural differences \citep{mcclintock1966cross}\footnote{A method taken even further using non 2×2 games e.g.~\cite{henrich2005economic}.}.


\begin{figure}[!t]
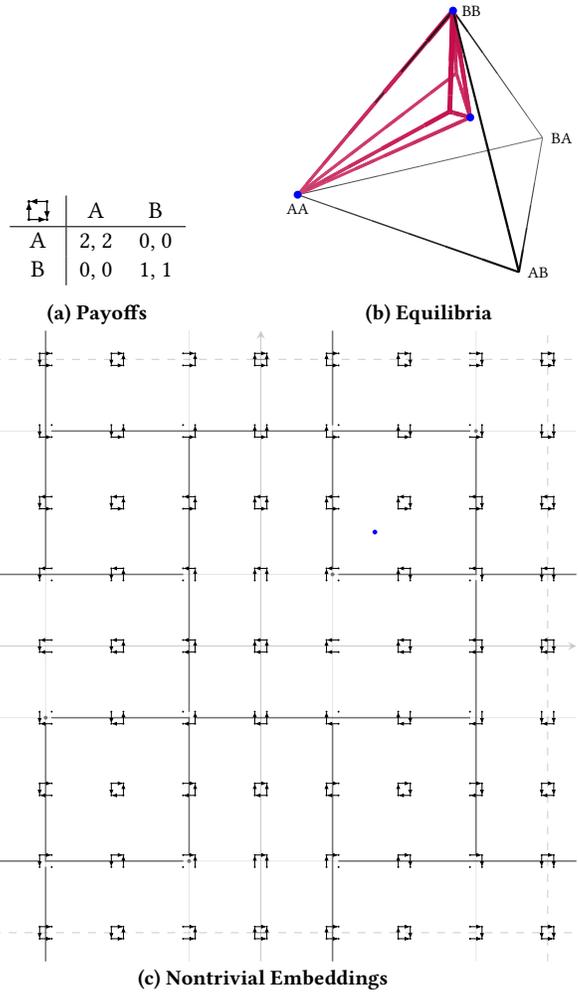

    \centering

    \begin{subfigure}[b]{0.48\linewidth}
        \centering
        \brpayoffstable{2}{0}{0}{1}{2}{0}{0}{1}
        \caption{Payoffs}
        \label{fig:exmple_visualizations_payoffs}
    \end{subfigure}\hfill
    \begin{subfigure}[b]{0.48\linewidth}
        \centering
        \polytope[width=\linewidth]{2}{0}{0}{1}{2}{0}{0}{1}
        \caption{Equilibria}
        \label{fig:exmple_visualizations_equilibria}
    \end{subfigure}

    \begin{subfigure}[b]{1.0\linewidth}
        \centering
        \nontrivialembedding[width=1.0\linewidth,no axes labels,no tick labels,no best-response names,points=data/coordinationpoint.dat]
        \vspace{-0.5em}
        \caption{Nontrivial Embeddings}
        \label{fig:exmple_visualizations_embedding}
    \end{subfigure}

    \caption{Example 2×2 visualizations for a coordination game produced by the \texttt{twoxtwogame} package.}
    \label{fig:example_visualizations}
\end{figure}

\begin{figure}[!t]

    \noindent\fbox{%
        \parbox{0.96\linewidth}{%
            The famous 2×2 normal-form game, \ordgraph{-1}{-3}{0}{-2}{-1}{0}{-3}{-2}~Prisoner's Dilemma, has a single pure Nash equilibrium (NE), where both players defect (\jmgraph{0}{0}{0}{1}). Traffic Lights~\ordgraph{0}{0}{0}{0}{0}{0}{0}{0} has two pure NE (\jmgraph{0}{1}{0}{0} and \jmgraph{0}{0}{1}{0}), one mixed NE (\jmgraph{1/100}{9/100}{9/100}{81/100}), and a more sensible correlated equilibrium solution (\jointgraph{0.0}{0.5}{0.5}{0.0}).
        }%
    }

    \caption{Example inline visualizations for 2×2 games, including payoffs and equilibria.}
    \label{fig:example_inline_visualizations}
\end{figure}

There is also extensive research concerned directly with the set of 2×2 games itself  \citep{rapoport1976_the_2x2_game_book}, including naming schemes \citep{bruns2015_names_for_games}, topologies \citep{robinsonandgoforth2005_topology_of_2x2_games_book}, taxonomies \citep{kilgour1988_taxonomy_of_all_ordinal_2x2_games,walliser1988_simplifed_taxonomy_2x2}, periodic tables \citep{goforth2005_periodic_table_of_games}, parameterizations \citep{harris1969_geometric_classification_of_symmetric_2x2}, embeddings \citep{marris2023_equilibrium_invariant_embedding_2x2_arxiv}, and other categorizations \citep{huertasrosero2003_cartography_of_symmetric_2x2,boors2022_2x2_game_classification_by_decomposition,germano2006_geometry_of_nf_games} of 2×2 games have been extensively studied. 2×2 games are utilized so frequently that their names have entered popular culture, for example: Chicken, Prisoner's Dilemma, Stag Hunt \citep{skyrms2004_stag_hunt}, Battle of the Sexes \citep{luce1957games}, and Matching Pennies.

2×2 games typically require 8 parameters to describe: each of the two players receives a payoff (equivalently a reward, or negative cost) for each possible joint action, of which there are $2 \times 2 = 4$. These can be easily represented as a table. However, this description is often clunky to insert inline into text, the game's properties are not immediately salient from raw numerical values, and naming such a game is not always easy. As a result, graphical representations have been developed for 2×2 ordinal games \citep{bruns2015_names_for_games} and 2×2 best-response dynamics \citep{marris2023_equilibrium_invariant_embedding_2x2_arxiv} (Section~\ref{sec:payoff_graphs}), which the \texttt{twoxtwogame} package implements. Representations of behaviour in 2×2 games can be described as joint of marginal distributions which can also be visualized inline (Section~\ref{sec:dist_graphs}). More complex is the set of equilibria, which are polytopes for correlated equilibria and points or lines for Nash equilibria, which can both be represented on a simplex (Section~\ref{sec:equilibria}). Finally, 2×2 games can be embedded in two dimensions \citep{marris2023_equilibrium_invariant_embedding_2x2_arxiv} allowing properties to be read from a point's position on a graph (Section~\ref{sec:embeddings}).

This work promotes a \LaTeX{} package, \texttt{twoxtwogame}, that has collections of tools and visualizations for 2×2 games. The package is developed on GitHub\footnote{\url{https://github.com/google-deepmind/twoxtwogame}} and distributed on Comprehensive TeX Archive Network (CTAN)\footnote{\url{https://www.ctan.org/pkg/twoxtwogame}}. The goal is for these visualizations to be flexible, well-tested, and of publication quality. To this end we build upon PGF/TikZ/PGFplots \citep{pgftikz2021,pgfplot2021} for producing crisp vector-graphics than can be inserted as figures or inline into \LaTeX{}. The figures produced in this paper are generated using this package. The \texttt{twoxtwogame} package is distributed with documentation that references all the commands available within it. Therefore the goal of this paper is not as a command reference for the package, but instead to promote and explain the package, and provide a mechanism for citing the package. Example figures produced by the package are shown in Figure~\ref{fig:example_visualizations} and Figure~\ref{fig:example_inline_visualizations}.

Furthermore, we hope that such a library will encourage some standardization of the language and visualizations we use to describe 2×2 games. Standard notation is important in many research fields. It is useful to succinctly and exactly communicate concepts. Convergence on standard ways to represent and name games is enabled by easy to use, free and open tools are available for the community to utilize and build upon. We hope the package serves the research community in this respect.

\section{Background and Notation}

A normal-form game (NFG) consists of $N$ players, $p \in \{1, ..., N\}$, of which each has $|\mathcal{A}_p|$ pure strategies, $a_p \in \{ a_p^1, ..., a_p^{|\mathcal{A}_p|} \}$. Players play a strategy simultaneously, resulting in a joint strategy, $a = (a_1, ..., a_N) \in \mathcal{A}_1 \times ... \times \mathcal{A}_N$. A joint strategy, $a \in \mathcal{A}$, results in a payoff for each player, defined by a payoff function, $G_p(a) = G_p(a_1, ..., a_N)$. In the most general case, the payoff function is a table (sometimes called a payoffs\footnote{We use plural, ``payoffs'' when referring to all players' payoffs, $(G_1, ..., G_N)$, and singular ``payoff'' when referring to a single player's payoff, $G_p$.} tensor, $G$) consisting of $N \prod_p |\mathcal{A}_p| = N |\mathcal{A}|$ entries. Sometimes we consider each player's payoff tensor separately, $G_p$, which each consisting of $\prod_p |\mathcal{A}_p|$ entries. An element of this tensor can be retrieved via the strategy parameters, $G_p(a_1^A, a_2^B, ...)$, or defined directly as a scalar, $g_p^{AB...}$.

Players need not play deterministically, they may randomize over their strategies. Doing so is called playing a mixed strategy and is defined by the marginal probability distribution, $\sigma_p(a_p) \in \Delta^{|\mathcal{A}_p| - 1}$, where $\Delta$ is a probability simplex. In this case a player receives a payoff, $g^{\sigma}_p = G_p(\sigma_1(a_1), ..., \sigma_N(a_N)) = \sum_{a \in \mathcal{A}} \sigma_p(a_1) ... \sigma_p(a_N) \allowbreak G_p(a_1, ..., a_N)$. Furthermore, players could coordinate their play, allowing them to play a joint mixed strategy, $\sigma(a) \in \Delta^{|\mathcal{A}| - 1}$. This results in a payoff $g^{\sigma}_p = G_p(\sigma(a)) = \sum_{a} \sigma(a) G_p(a)$. Note that mixed strategies are a subset of joint mixed strategies because $\sigma(a)= \otimes_p \sigma(a_p)$.

\subsection{Properties of Normal-Form Games}

Researchers are interested in studying how rational payoff-maximizing players behave in NFGs. One concept is the best-response which is the set of best strategies a player should take in response to other players' strategies.
\begin{align} \label{eq:br}
    \mathcal{A}^\text{BR}_p = \arg \max_{a'_p} \sum_{a_{-p}} \sigma(a_{-p}) G_p(a'_p, a_{-p})
\end{align}

As well as best-responses, researchers are interested in solutions that are stable: ones where no player has any incentive to unilaterally deviate. There are a number of different ways of defining stability. Often such classes of solutions are known as equilibria. The simplest definition of an equilibrium is the coarse correlated equilibrium (CCE) \citep{hannan1957_cce,moulin1978_cce} which is defined for coordinated play using a set of linear inequality constraints, $\forall p \in [1, N], a'_p \in \mathcal{A}_p$, where $a'_p$ are candidate deviation strategies.
\begin{align} \label{eq:cce}
    \sum_{a} \sigma(a) \left( G_p(a'_p, a_{-p}) - G_p(a) \right) \leq 0
\end{align}
A related concept call correlated equilibrium (CE) is equivalent to CCEs in games with only two strategies. 

The most well known solution concept for decentralized play is the Nash equilibrium (NE) \citep{nash1951_neq}. It is the same as CCE, with the additional constraint that the joint must factorize. 
\begin{align} \label{eq:ne}
    \sum_{a} \left(\otimes_p \sigma_p(a_p) \right) \left( G_p(a'_p, a_{-p}) - G_p(a) \right) \leq 0
\end{align}

These constraints always result in a nonempty set of equilibria. These sets are invariant to certain transformations of the payoffs, including affine transformations. Additionally, permutation of strategies, and players results in equivalent games. Exploiting these invariances allows defining game embeddings \citep{marris2023_equilibrium_invariant_embedding_2x2_arxiv}.

\subsection{2×2 Games}

A 2×2 game is an NFG with two players ($N=2$, $p = \{1, 2\}$), each with two actions ($|\mathcal{A}_1| = |\mathcal{A}_2| = 2$). A 2×2 game has a payoffs tensor, $G = (G_1, G_2)$, with $4 + 4 = 8$ entries.

A subclass of commonly studied 2×2 games are ordinal games \citep{rapoportandguyer1966_taxonomy_of_2x2_games} which have payoff tensors consisting of only integers, $\{1, 2, 3, 4\}$, each of which occurs only once, $(g_p^{AA}, g_p^{AB}, g_p^{BA}, g_p^{BB}) = \text{perm}[\allowbreak (1, 2, 3, 4)]$. In a partial ordinal game entries can be repeated, $G_p(a) \in \{1, 2, 3, 4\}$. There are $4! \times 4! = 576$ ordinal games, $144$ up to strategy permutation invariance, and $78$ up to additionally player invariance. There are $726$ partial ordinal games \citep{fraser1986_non_strict_2x2_games}, up to strategy and player invariances. Any 2×2 game can be mapped to a partial ordinal game by considering the order of the elements in the payoff. Therefore partial ordinal games can also be considered ``classes'' of games.

Another equivalence class of 2×2 games, introduced in \cite{borm1987_classification_of_2x2_games} and \cite{marris2023_equilibrium_invariant_embedding_2x2_arxiv}, is defined in terms of equilibria and best-response dynamics of games. This results in 81 classes of games, or 15 up to permutation invariance. Additionally, the embeddings of these games can be visualized in two dimensions.

\section{Payoff Graphs}
\label{sec:payoff_graphs}

The \texttt{twoxtwogame} package contains commands that produce graph representation of 2×2 games. These most commonly operate directly on the elements in the payoff. The arguments to the commands are on the flattened payoffs: starting with the row player, row-major, then for the column player, row-major (Figure~\ref{fig:payoffs_arguments}).
\begin{figure}
    \centering
    \begin{subfigure}[t]{0.49\linewidth}
        \centering
        \payoffstable[label=$G$]{a}{b}{c}{d}{e}{f}{g}{h}
        \caption{Payoffs}
    \end{subfigure}\hfill
    \begin{subfigure}[t]{0.49\linewidth}
        \centering
        \jointtable[label=$\sigma$]{a}{b}{c}{d}
        \caption{Joint}
    \end{subfigure}
    \cprotect\caption{Arguments for payoffs and joint commands. Typically a command is called with \texttt{\textbackslash payoffscommand\{a\}\{b\}\allowbreak \{c\}\{d\}\allowbreak \{e\}\{f\}\{g\}\{h\}} or \texttt{\textbackslash jointcommand\{a\}\{b\}\allowbreak \{c\}\{d\}}.}
    \label{fig:payoffs_arguments}
\end{figure}

\subsection{Partial-Ordinal Graphs}

Visualizing an ordinal game as a graph is simple: with the elements of the payoff matrix as nodes, trace edges from the lowest to highest payoff to form a directed graph. This can be done separately for each player (Figure~\ref{fig:ordinal_graph_example}). This approach has appeared previously in the literature \citep{bruns2015_names_for_games}.

\begin{figure}
    \centering
    \begin{subfigure}[t]{0.49\linewidth}
        \centering
        \payoffstable{1}{2}{3}{4}{1}{3}{4}{2} ~~~~~~
        \begin{tabular}{c|cc}
             & A & B  \\ \hline
            A & \multicolumn{2}{c}{\multirow{3}{*}{\raisebox{1.2em}{\huge{\ordgraph{1}{2}{3}{4}{1}{3}{4}{2}}}}} \\
            B &
        \end{tabular}
        \caption{\centering Ordinal}
        \label{fig:ordinal_graph_example}
    \end{subfigure}\hfill
    \begin{subfigure}[t]{0.49\linewidth}
        \centering
        \payoffstable{1}{1}{2}{2}{1}{2}{2}{3} ~~~~~~
        \begin{tabular}{c|cc}
             & A & B  \\ \hline
            A & \multicolumn{2}{c}{\multirow{3}{*}{\raisebox{1.2em}{\huge{\ordgraph{1}{1}{2}{2}{1}{2}{2}{3}}}}} \\
            B &
        \end{tabular}
        \caption{\centering Partial Ordinal}
        \label{fig:partial_ordinal_graph_example}
    \end{subfigure}

    \caption{Example graphs for (partial) ordinal games. The row player's preference order is given by the black arrow and the column player's preference is given by the gray arrow.}
    \label{fig:ordinal_graph_examples}
\end{figure}

For partial ordinal games, where elements of the payoff may have equal values, the graphing rules are more complex. In this case, split elements of the payoff into sets with equal payoff. Then all elements of each set point to all elements of the next set with increasing payoff (Figure~\ref{fig:partial_ordinal_graph_example}). This approach is invented by this package.

Any NFG can be converted to a partially ordinal game by considering the partial order of the payoff elements. The resulting game retains many of the properties of the original game. Therefore NFGs with arbitrary payoffs can be sensibly visualized with such graphs, and properties such as equilibria, dominant strategies, player preferences can be directly and intuitively read from them.

These graphs can be produced by the \texttt{twoxtwogame} package using the \verb!\ordgraph! command which takes 8 numerical arguments. Importantly, this command can be inserted inline to text. For example, we can use this command to prefix common 2×2 games: \ordgraph{1}{-1}{-1}{1}{-1}{1}{1}{-1}~Matching Pennies, \ordgraph{1}{-1}{-1}{1}{1}{-1}{-1}{1}~Coordination, or \ordgraph{1}{3}{0}{2}{1}{0}{3}{2}~Prisoner's Dilemma. See Table~\ref{tab:ordgraph_command_examples} for further example usage. A variety of optional arguments control the colors, offsets, and styles of the graphs (described in the documentation bundled with the package).

\begin{table}[t!]
    \cprotect\caption{Examples for representing the order of payoffs in 2×2 games using the \texttt{\textbackslash ordgraph} command.}
    \label{tab:ordgraph_command_examples}
    \begin{tabular}{lr}
        Command & Output \\ \hline
        \verb!\ordgraph{1}{-1}{-1}{1}{-1}{1}{1}{-1}! & \ordgraph{1}{-1}{-1}{1}{-1}{1}{1}{-1} \\
        \verb!\ordgraph{1}{-1}{-1}{1}{1}{-1}{-1}{1}! & \ordgraph{1}{-1}{-1}{1}{1}{-1}{-1}{1} \\
        \verb!\ordgraph{1}{3}{0}{2}{1}{0}{3}{2}! & \ordgraph{1}{3}{0}{2}{1}{0}{3}{2}  \\
        \verb!\ordgraph{.9}{.5}{.1}{.9}{.2}{.7}{.6}{1} ! & \ordgraph{.9}{.5}{.1}{.9}{.2}{.7}{.6}{1}
    \end{tabular}
\end{table}

\subsection{Best-Response Graphs}

A similar graph representation can be produced by just considering the best-responses to the other player's strategy \citep{marris2023_equilibrium_invariant_embedding_2x2_arxiv}. A directed edge indicates the preferred strategy a player should choose. If a player is indifferent between two strategies, the edge is omitted. Vertical edges correspond to the row player's preference in response fixed column player strategies, and the horizontal edges correspond to  column player's preference in response to fixed row player strategies (Figure~\ref{fig:br_graph_examples}).

\begin{figure}[t]
    \centering
    \begin{subfigure}[t]{0.49\linewidth}
        \centering
        \payoffstable{1}{2}{3}{4}{1}{3}{4}{2} ~~~~~~
        \begin{tabular}{c|cc}
             & A & B  \\ \hline
            A & \multicolumn{2}{c}{\multirow{3}{*}{\raisebox{1.2em}{\huge{\brgraph{1}{2}{3}{4}{1}{3}{4}{2}}}}} \\
            B &
        \end{tabular}
        \caption{\centering}
        \label{fig:br_graph_example}
    \end{subfigure}\hfill
    \begin{subfigure}[t]{0.49\linewidth}
        \centering
        \payoffstable{1}{1}{2}{2}{1}{2}{2}{3} ~~~~~~
        \begin{tabular}{c|cc}
             & A & B  \\ \hline
            A & \multicolumn{2}{c}{\multirow{3}{*}{\raisebox{1.2em}{\huge{\brgraph{1}{1}{2}{2}{1}{2}{2}{3}}}}} \\
            B &
        \end{tabular}
        \caption{\centering}
        \label{fig:partial_br_graph_example}
    \end{subfigure}
    \caption{Example best-response graphs \citep{marris2023_equilibrium_invariant_embedding_2x2_arxiv}. No colour is necessary to disambiguate the players because vertical edges are the row player and horizontal edges are the column player.}
    \label{fig:br_graph_examples}
\end{figure}

These graphs can be produced by the \texttt{twoxtwogame} package using the \verb!\brgraph! command which takes 8 numerical arguments.  For example, we can use this command to prefix common 2×2 games: \brgraph{1}{-1}{-1}{1}{-1}{1}{1}{-1}~Matching Pennies, \brgraph{1}{-1}{-1}{1}{1}{-1}{-1}{1}~Coordination, or \brgraph{1}{3}{0}{2}{1}{0}{3}{2}~Prisoner's Dilemma. See Table~\ref{tab:brgraph_command_examples} for further example usage. A variety of optional arguments control the colors, offsets, and styles of the graphs (described in the package documentation).

\begin{table}[t!]
    \cprotect\caption{Examples of representing the best-response dynamics of 2×2 games using the \texttt{\textbackslash brgraph} command.}
    \label{tab:brgraph_command_examples}
    \begin{tabular}{lr}
        Command & Output \\ \hline
        \verb!\brgraph{1}{-1}{-1}{1}{-1}{1}{1}{-1}! & \brgraph{1}{-1}{-1}{1}{-1}{1}{1}{-1} \\
        \verb!\brgraph{1}{-1}{-1}{1}{1}{-1}{-1}{1}! & \brgraph{1}{-1}{-1}{1}{1}{-1}{-1}{1} \\
        \verb!\brgraph{1}{3}{0}{2}{1}{0}{3}{2}! & \brgraph{1}{3}{0}{2}{1}{0}{3}{2}  \\
        \verb!\brgraph{.9}{.5}{.1}{.9}{.2}{.7}{.6}{1} ! & \brgraph{.9}{.5}{.1}{.9}{.2}{.7}{.6}{1} 
    \end{tabular}
\end{table}

There are only a finite number of such graphs, and after removing redundant graphs that are equivalent under permutation, it results in 15 equivalence classes. Each of these equivalence classes is named in \cite{marris2023_equilibrium_invariant_embedding_2x2_arxiv}. The \texttt{twoxtwogame} package provides a command for retrieving the game of any 2×2 games (Table~\ref{tab:brname_command_examples}).

\begin{table}[t!]
    \cprotect\caption{Examples of finding best-response names of 2×2 games using the \texttt{\textbackslash brname} command.}
    \label{tab:brname_command_examples}
    \begin{tabular}{lr}
        Command & Output \\ \hline
        \verb!\brname{1}{-1}{-1}{1}{-1}{1}{1}{-1}! & \brname{1}{-1}{-1}{1}{-1}{1}{1}{-1} \\
        \verb!\brname{1}{-1}{-1}{1}{1}{-1}{-1}{1}! & \brname{1}{-1}{-1}{1}{1}{-1}{-1}{1} \\
        \verb!\brname{1}{3}{0}{2}{1}{0}{3}{2}! & \brname{1}{3}{0}{2}{1}{0}{3}{2}  \\
        \verb!\brname{.9}{.5}{.1}{.9}{.2}{.7}{.6}{1} ! & \brname{.9}{.5}{.1}{.9}{.2}{.7}{.6}{1}
    \end{tabular}
\end{table}

\section{Distribution Graphs}
\label{sec:dist_graphs}

Most commonly, the behaviour of players if the goal of studying games. This behaviour may be stochastic and therefore is best described through probability distributions. If behaviour is uncoordinated (like in a Nash equilibrium), marginal distributions are the best way to describe actions of players. However if play is coordinated joint or conditional distributions may be a better option. These can be represented in a table, however shaded visualizations of distributions are a useful and succinct way of representing player behaviour (Table~\ref{tab:dist_command_examples}). The \texttt{twoxtwogame} package provides commands for visualizing marginal, joint, and conditional distributions.

\begin{table}[h!]
    \caption{Examples of inline distribution commands used to visualize 2×2 distributions.}
    \label{tab:dist_command_examples}
    \begin{tabular}{lr}
        Command & Output \\ \hline
        \verb!\jointgraph{.4}{.3}{.1}{.2}! & \jointgraph{.4}{.3}{.1}{.2} \\
        \verb!\rowcondgraph{.4}{.3}{.1}{.2}! & \rowcondgraph{.4}{.3}{.1}{.2} \\
        \verb!\colcondgraph{.4}{.3}{.1}{.2}! & \colcondgraph{.4}{.3}{.1}{.2} \\
        \verb!\marginalgraph{.4}{.3}{.1}{.2}! & \marginalgraph{.4}{.3}{.1}{.2} \\
        \verb!\jmgraph{.4}{.3}{.1}{.2} ! & \jmgraph{.4}{.3}{.1}{.2}
    \end{tabular}
\end{table}

\section{Equilibria}
\label{sec:equilibria}

There are many solution concepts for games, but perhaps the most ubiquitous are equilibria concepts: distributions of play where no player is incentivized to unilaterally deviate. These solutions are guaranteed to exist, but there may be many for each game.

The \texttt{twoxtwogame} package provides vector graphic visualizations of the set of equilibria of any 2×2 game. In order to visualize the equilibria of 2×2 games we need to visualize the mixed joint strategy space, $\sigma(a) \in \Delta^3$, of 2×2 games. This lives on the 3-simplex which can be visualized using four points in 3D space. Choose equally spaced points for each of the four pure joint strategies. Then any convex combination of these four points represents all valid mixed joint strategies. This results in a tetrahedron of valid points in Cartesian coordinates.

\begin{figure*}
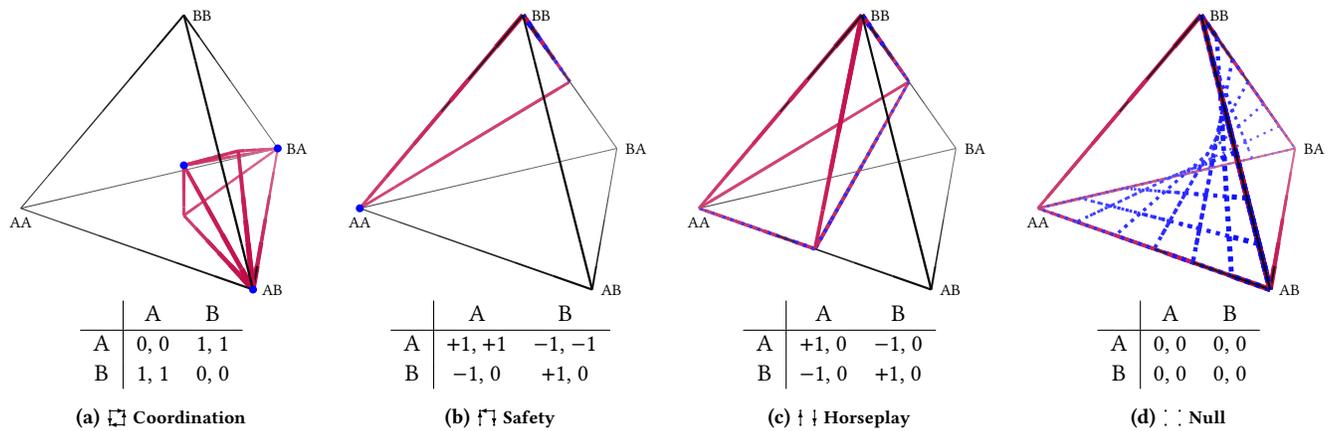

    \centering

    \begin{subfigure}[t]{0.24\linewidth}
        \centering
        \polytope[width=\linewidth]{0}{1}{1}{0}{0}{1}{1}{0}
        \vspace{-0.2em}
        
        \payoffstable{0}{1}{1}{0}{0}{1}{1}{0}
        \caption{\footnotesize\brgraph{0}{1}{1}{0}{0}{1}{1}{0}~\brname{0}{1}{1}{0}{0}{1}{1}{0}}
        \label{fig:simplex_coordination}
    \end{subfigure}\hfill
    \begin{subfigure}[t]{0.24\linewidth}
        \centering
        \polytope[width=\linewidth]{+1}{-1}{-1}{+1}{+1}{-1}{0}{0}
        \vspace{-0.2em}
        
        \payoffstable{+1}{-1}{-1}{+1}{+1}{-1}{0}{0}
        \caption{\footnotesize\brgraph{+1}{-1}{-1}{+1}{+1}{-1}{0}{0}~\brname{+1}{-1}{-1}{+1}{+1}{-1}{0}{0}}
        \label{fig:simplex_safety}
    \end{subfigure}\hfill
    \begin{subfigure}[t]{0.24\linewidth}
        \centering
        \polytope[width=\linewidth]{+1}{-1}{-1}{+1}{0}{0}{0}{0}
        \vspace{-0.2em}
        
        \payoffstable{+1}{-1}{-1}{+1}{0}{0}{0}{0}
        \caption{\footnotesize\brgraph{+1}{-1}{-1}{+1}{0}{0}{0}{0}~\brname{+1}{-1}{-1}{+1}{0}{0}{0}{0}}
        \label{fig:simplex_horseplay}
    \end{subfigure}\hfill
    \begin{subfigure}[t]{0.24\linewidth}
        \centering
        \polytope[width=\linewidth]{0}{0}{0}{0}{0}{0}{0}{0}
        \vspace{-0.2em}
        
        \payoffstable{0}{0}{0}{0}{0}{0}{0}{0}
        \caption{\footnotesize\brgraph{0}{0}{0}{0}{0}{0}{0}{0}~\brname{0}{0}{0}{0}{0}{0}{0}{0}}
        \label{fig:simplex_zero}
    \end{subfigure}\hfill
    \hspace{0.24\linewidth}
    
    \caption{(C)CE polytopes (purple) and NE (blue) in four example 2×2 games. The visualization uses native \LaTeX{} libraries and therefore is high resolution and respects the style of the surrounding document.}
    \label{fig:example_equilibria}
\end{figure*}

(Coarse) Correlated Equilibria ((C)CEs) are convex polytopes within the simplex. The polytopes are defined directly from the definition of the (C)CE (Equation~\eqref{eq:cce}), and the edges of the polytope are represented by lines in the visualization. Nash equilibria (NEs) of a game are either isolated points, a continuum of dashed lines, or a manifold. NEs are a subset of (C)CEs, whose joints factorize, so should appear within or on the boundary of the (C)CE polytope. Figure~\ref{fig:example_equilibria} shows the set of equilibria of four example 2×2 games.

These figures are produced directly in \LaTeX{} which means a) the graphs use vector graphics so are natively high resolution, and b) the font style and size is shared with the surrounding document. Furthermore, care has also been taken to use perspective to aid comprehension.

\section{Embeddings}
\label{sec:embeddings}

\begin{figure}[!t]
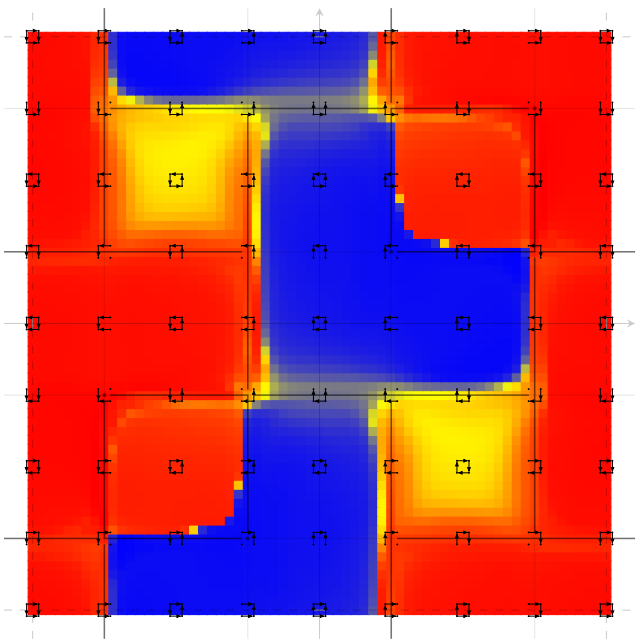

    \centering
    \nontrivialembedding[width=1.0\linewidth,no axes labels,no tick labels,no best-response names,matrix=data/embeddingheatmap.dat]
    \caption{Example heatmap embedding visualizations of nontrivial games. Data from \cite{liu2024_nfgtransformer}.}
    \label{fig:example_nontrivial_embedding_heatmap}
\end{figure}

Previous work \cite{marris2023_equilibrium_invariant_embedding_2x2_arxiv} introduced a game theoretic equilibrium-\allowbreak invariant embedding: a lower dimensional description of a normal-form game that preserves its set of equilibria. For the particular case of 2×2 games, only two variables are required, and therefore can be visualized in two dimensions. A game can be plotted on this space with a single point, or many games can be plotted as a point cloud. Properties of a game may be represented with a heatmap over the space of 2×2 games (for example, Figure~\ref{fig:example_nontrivial_embedding_heatmap}). These visualizations can be produced using the \texttt{\textbackslash nontrivial\allowbreak embedding} command.

\section{Contributing}

The project is maintained on GitHub\footnote{\url{https://github.com/google-deepmind/twoxtwogame}} under the Apache 2 license and contributions are welcome. If you would like to contact us regarding anything related to \texttt{twoxtwogame}, please create an issue on GitHub so that the team is notified. It is especially important to communicate with the maintainers before embarking on large contributions to ensure your time is used most productively.

\section{Conclusion}

The \texttt{twoxtwogame} \LaTeX{} package is a collection of commands for visualizing 2×2 normal-form games. We hope that the provision of professional and standardized visualizations will help researchers communicate their findings more succinctly and precisely. Furthermore, we hope such tools will help standardize terminology.


\bibliographystyle{ACM-Reference-Format} 
\bibliography{bibtex,bibtex_colab}

\clearpage
\appendix


\end{document}